\documentstyle[eqsecnum,aps,epsfig]{revtex}

\begin{document}
\draft
\preprint{HEP/123-qed}
\title{Potentials between static SU(3) sources in the
fat-center-vortices model}
\author{Sedigheh Deldar}
\address{
Washington University, St. Louis, MO, 63130
}
\date{\today}
\maketitle

\begin{abstract}

The potentials between static sources in various
representations in $SU(3)$ are calculated based on the
fat-center-vortices model of Faber, Greensite and Olejn\'{i}k. 
At intermediate distances, most distributions of the flux within vortices lead 
to potentials that are
qualitatively in agreement with ``Casimir scaling,'' which says that
the string tension is proportional to the quadratic operator of the 
representation. However, at the quantitative level, violations of Casimir
scaling are generally much larger than those seen in numerical simulations,
indicating that additional physical input to the fat-center-vortices model
is required. At large distances, screening occurs for zero-triality 
representations; for the representations with non-zero triality 
the string tension equals that of the fundamental representation.
Some rather ``unphysical'' flux distributions can lead to violations of Casimir
scaling at intermediate distances and violations of the expected ordering
of representations at large distances.
\end{abstract}

\maketitle

\section{INTRODUCTION}

A variety of models have been introduced in an attempt to explain
quark confinement. The center vortices model
\cite{Hoof79,Corn79} introduced in the late 1970's
by 't Hooft is one of those attempts.
A center vortex is a topological line-like (in D=3 dimensions) or 
surface-like (in D=4 dimensions) field configuration. 
The vortex carries magnetic flux quantized
in terms of elements of the center of the group. The fluxes form narrow
tubes with constant energy per unit length (surface).
In order for the vortex to have finite energy per unit length, the
gauge potential at large transverse
distances must be a pure gauge. However, the gauge transformation
which produces that potential is non-trivial. It is discontinuous
by an element of the gauge center.
It is the non-trivial nature of the gauge transformation which forces
the vortex core to have non-zero energy and makes
the vortex topologically stable.
Faber, Greensite and Olejn\'{i}k \cite{Fabe97} introduced
fat-center-vortices to obtain confinement of both
fundamental and higher representation static sources.
According to the fat-center-vortices model, the vacuum is a condensate
of vortices of some finite thickness. Confinement is produced by the
independent fluctuations of the vortices piercing each unit area of a
Wilson loop.

Faber {\it et al.} explicitly worked out the model for SU(2) and gave results
for a particular flux distribution within the vortices. Here, I work in
SU(3) and investigate a wide variety of vortex flux distributions. I study the 
existence of Casimir scaling at intermediate distances and the pattern of
screening at large distances. 
The model confirms Casimir scaling qualitatively at intermediate distances, 
however, in general the model 
does not agree with Casimir scaling  quantitatively, in contrast to the precise
agreement recently found by numerical simulations \cite{Deld98,Bali99}.
The lack of convexity of the potentials predicted by the model is also 
discussed.
For completeness, I first briefly explain their model and then apply it to 
SU(3), using it to study the potentials between static quarks for the 
fundamental and several other representations.

\section{The model of Faber, Greensite and Olejn\'{i}k}

In the fundamental representation of $SU(N)$, a center vortex linked to a
Wilson loop has the effect of multiplying the Wilson loop
by the gauge group center,
\begin{equation}
W(C) \rightarrow \exp^{\frac{2\pi i n}{N}} W(C)~~~~~~~~n=1,2,...,N-1.
\end{equation}

Based on the vortex theory, the area law for a Wilson loop is due to
the quantum fluctuations in the number of center vortices linking the loop.
Adjoint Wilson loops are not affected by center vortices unless the
vortex core overlaps the perimeter of the
loops. If the vortex thickness is large enough, the fat-center-vortices model 
can explain confinement and the Casimir scaling of higher representation 
string tensions. The average Wilson loop
predicted by this model has the following form:
\begin{equation}
<W(C)> = \prod_{x_{v}} \{ 1 - \sum^{N-1}_{n=1} f_{n} (1 - Re {\cal G}_{r}
          [\vec{\alpha}^n_{C}(x_{v})])\},
\label{sigmac}
\end{equation}
where $x_{v}$ is the location of the center of the vortex, $A$ is the area of 
the loop $C$, and ${\cal G}_{r}$ is defined as:
\begin{equation}
{\cal G}_{r}[\vec{\alpha}] = \frac{1}{d_{r}} Tr \exp[i\vec{\alpha} . \vec{H}].
\label{gr}
\end{equation}
$d_{r}$ is the dimension of representation r, and
$\{H_{i}\}$ is the subset of the generators needed to generate all
elements of the center of the group.\@ (For $SU(3)$, $\lambda_{8}$
is sufficient.)
The parameter $f$ represents the probability that any given unit area is
``pierced" by a vortex; {\it {i.e.}}, a line running through the center of
the vortex tube intersects the area.

The parameter $\alpha_{c}(x)$ is determined by the
fraction of the vortex flux that is enclosed by the Wilson loop. Therefore 
$\alpha_{c}(x)$ depends on the profile of the vortex as well as the shape of 
the loop and the position $x_{v}$ of the center of the vortex 
relative to the perimeter. For SU(3) $\alpha_{c}(x)$ is equal to 
$4\pi/\sqrt{3}$, if the flux is entirely inside the minimal area of the loop 
and it is zero if the flux is entirely outside the minimal area of the loop.

\section{General features of Casimir scaling and screening}

Numerical simulations \cite{Camp86,Mich92,Mich98,Deld98,Bali99}
have shown that the potentials in $SU(3)$ at zero temperature
are linear at intermediate distances and roughly proportional to the Casimir 
operator. The proportionality of potential to the Casimir operator is known
as ``Casimir scaling.'' Assuming such scaling, the potentials of each 
representation to that of fundamental representation would be 2.5, 2.25,
4.5, 7, 4 and 6 for representations 6, 8, 10, 15-symmetric, 15-antisymmetric
and 27, respectively. The Casimir scaling regime is expected to exist for 
intermediate distances, perhaps extending from the onset of confinement to 
the onset of screening. 

Screening can be understood as follows:
Each representation can be labeled by the ordered pair $(n,m)$, with
n and m the original number of 3 and $\bar{3}$ which participated in
constructing the representation. Triality is defined as (n-m) mod 3.
Screening occurs for representations with zero triality:
$8 \equiv (1,1)$, $10 \equiv (3,0)$, and $27 \equiv (2,2)$.
For these representations, as the distance between the two adjoint
sources increases, the potential energy of the flux tube rises. A pair of
gluons pops of vacuum when this energy is equal or greater than the
twice of glue-lump mass. (A glue-lump is the ground state hadron with a
gluon field around a static adjoint source.) At large distances, the
static sources combine with the adjoint(8) charges (dynamic gluons) popped out
of the vacuum and produce singlets which screen.
Therefore the potential between static sources is no longer
$R$ dependent. Static sources in representations 10 and 27 transform into the 8
first by combining with a dynamic gluon, and then the 8 transforms into the 
singlet by combining with a second gluon.
\begin{equation}
8 \otimes 8= 27 \oplus \bar{10} \oplus 10 \oplus 8 \oplus 1,
\end{equation}
\begin{equation}
10\otimes 8= 8 \oplus 10 \oplus 27 \oplus 35,
\end{equation}
\begin{equation}
27 \otimes 8 = 64 \oplus 27 \oplus 27 \oplus 35 \oplus \bar{35} \oplus 10
 \oplus \bar{10} \oplus 8.
\end{equation}
Therefore, we expect the potential in representations 10 and 27 to screen only
at higher energy than in representation 8.
Static sources in representations with non-zero triality, $6 \equiv (2,0)$, $15_{s} \equiv (4,0) $
and $15_{a} \equiv (2,1)$, transform into the lowest order representation (3
and $\bar{3}$) by binding to the gluonic $8's$ which are popped out of the
vacuum:
\begin{equation}
6 \otimes 8 = \bar{3} \oplus  6 \oplus 15 \oplus 24,
\end{equation}
\begin{equation}
15_{a} \otimes 8= 42 \oplus \bar{24} \oplus 15_{a} \oplus 15_{a}
\oplus \bar{6} \oplus 3  \oplus 15_{s},
\end{equation}
\begin{equation}
15_{s} \otimes 8 = 48 \oplus 42 \oplus 15_{s} \oplus 15_{a}.
\end{equation}
15-symmetric changes to 15-antisymmetric first, so it needs to interact
with the 8's (popped from the vacuum) twice to transform to 3. Screening
does not occur for representations with non-zero triality,
since there is no way to get a zero
triality representation by crossing a non-zero one with any number of 8's.
As a result, the slope of the linear potentials of the representations
with non-zero triality changes to the slope of the fundamental one,
and a universal string tension is observed for large R. We expect the 
representation 15-symmetric to require a larger value of $R$ to approach the 
fundamental slope than representations $6$ or 15-antisymmetric 
because two pairs of 8's must be popped from the vacuum in the 15-symmetric 
case.

\section{Applying the Fat-center-vortices model to $SU(3)$}

To find the potential $V_{r}(R)$ in SU(3), first I need to find $H_{i}$ in
Eqn.\ \ref{gr} for each representation: 3, 6, 8, 10, 15-symmetric,
15-antisymmetric, and 27. For the fundamental representation,
$H_{1}=T_{8}=\frac{\lambda_{8}}{2}$; where $\lambda_{8}$ is the diagonal 
Gell-Mann matrix.

I obtain $T^r_{8}$ of other representations by using the tensor method.
Define $\{X^i_{r}; i=1,...,d_{r}\}$, which are the basis vectors for
the space on which the representation act. The corresponding
generators are obtained from \cite{Geor92}:
\begin{equation}
[T_{a}^{D_{1}\otimes D_{2}}]_{ix,jy}=[T_{a}^{D_{1}}]_{ij}\delta_{xy}+
       \delta_{ij}[T_{a}^{D_{2}}]_{xy}.
\label{prod_rep}
\end{equation}
$T_{a}'s$ are the group generators for representations $D_{1}$, $D_{2}$,
$D_{1}\otimes D_{2}$. The elements of $T^r_{8}$ can be found by:
\begin{equation}
T^r_{8} X^i_{r} = \sum^{d_{r}}_{j=1} C_{ij}X^j_{r}.
\label{Tr8}
\end{equation}

To study potentials for $SU(3)$, one needs to define an appropriate form of
the function $\alpha_{c}(x)$ in Eqn.\ \ref{sigmac}. To understand more about the
effect of the vortex profile on potentials and Casimir scaling, I assume a
density of flux $\rho(r)$ in an axially symmetric vortex core, where $r$ is
the radial distance from the vortex center. Let $\rho(r)=0$ for $r>a$
so the vortex has a sharp boundary at $r=0$.
Now let $\beta(x_{v})$ denote the amount of the flux of the vortex contained
in the region $x>0$. Thus, $\beta(x_{v})=0$ for $x_{v}< -a$, and
$\beta(x_{v})=\frac{4\pi}{\sqrt{3}}$ for $x_{v}>a$.  For $-a \leq x_{v} \leq a$, 
$\beta(x_{v})$ is determined by the integral of $\rho(r)$ over the
fraction of the vortex in the region $x>0$. Finally, let the Wilson loop
have sides $x=0$ and $x=R$. Then $\alpha_{c}(x_{v})$, the fraction of flux 
within the loop, is given by:
\begin{equation} 
\alpha_{R}(x_{v})= \beta(x_{v})-\beta(x_{v}-R), 
\label{alpha-def} 
\end{equation} 

\begin{figure}[htb]
\epsfxsize=.8 \hsize
\epsffile{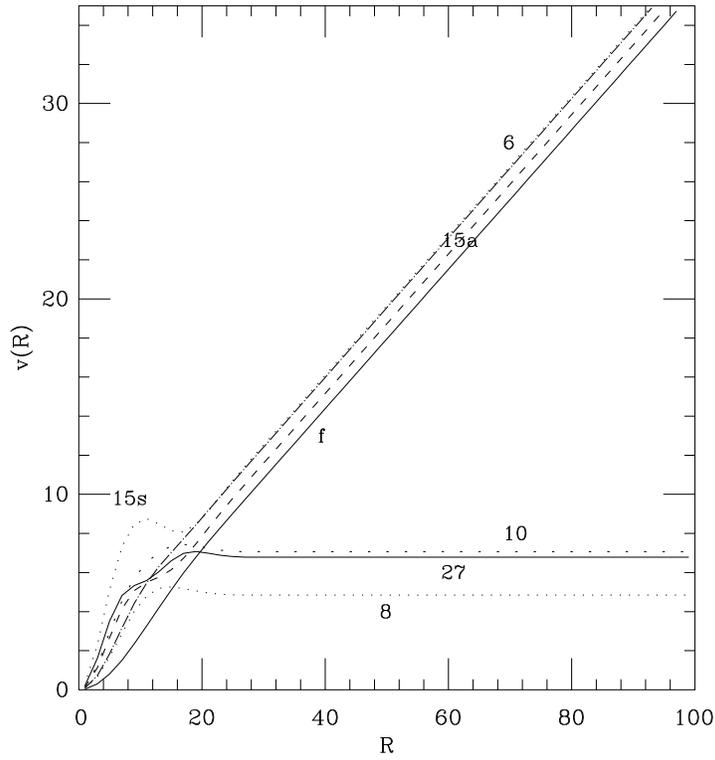}
\caption{Potentials for different representations using density distribution 
proportional to $\rho_{0}\exp(\frac{-.1}{(\frac{|r|}{a}-1.0)^2})$ for $-a<x<a$.
The fundamental representation is shown by the letter ``f''. At intermediate
distances ($3<R<5$), the ratios of the string tension of representations 8, 6, 
15-antisymmetric, 10, 27 and 15-symmetric to the one of the fundamental 
representation are 2.02, 2.21, 3.1, 3.4, 3.8 and 5.6, respectively. This is 
qualitatively in agreement with the Casimir ratios which are 2.25, 2.5, 4, 
4.5, 6 and 7, respectively.}
\label{expnroa}
\end{figure}

A simple choice for $\rho(r)$ that I have tried is a uniform distribution 
$\rho(r)=\rho_{0}$ for $r<a$. Another possibility with a smoother edge at
$r=a$, is:
\begin{equation}
\rho(r)=\rho_{0}\exp[{\frac{-b}{(\frac{|r|}{a}-1)^2}}],
\end{equation}
where $b$ is an adjustable constant and $\rho_{0}$ is fixed by the requirement 
that the total flux is $\frac{4\pi}{\sqrt{3}}$.
Fig.\ \ref{expnroa} shows potentials obtained from this flux distribution 
with $b=0.1$, $a=20$ and $f=0.1$.
From the plot, it can be seen that, for each representation, there exists a
region in which the potential is approximately linear and qualitatively in
agreement with Casimir scaling. Screening
occurs for representations 8, 10 and 27 while the slope of the potentials
for representations 6, 15-symmetric and 15-antisymmetric changes to the
slope of the fundamental representation. 
Note the non-convexity near $R=8$ to $R=20$ for all representations, and
especially representations 15-symmetric and 27.
Even though the fat-center-vortices model predicts
some of the expected behavior of the potential between static quarks,
it has some limitations. In particular, it violates the fact that the
potential should be always a convex function of distance \cite{Bach86}.

\begin{figure}[htb]
\epsfxsize=.8 \hsize
\epsffile{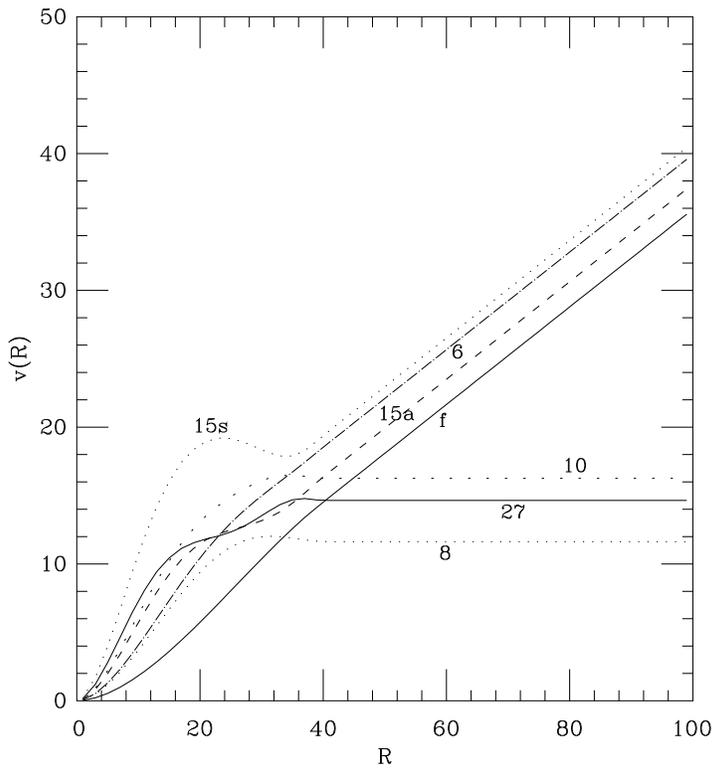}
\caption{Same as Fig.\ \ref{expnroa} but for density distribution equal to zero 
everywhere except at the vortex circumference($r=a$).}
\label{Pdens}
\end{figure}

\begin{figure}[htb]
\epsfxsize=.8 \hsize
\epsffile{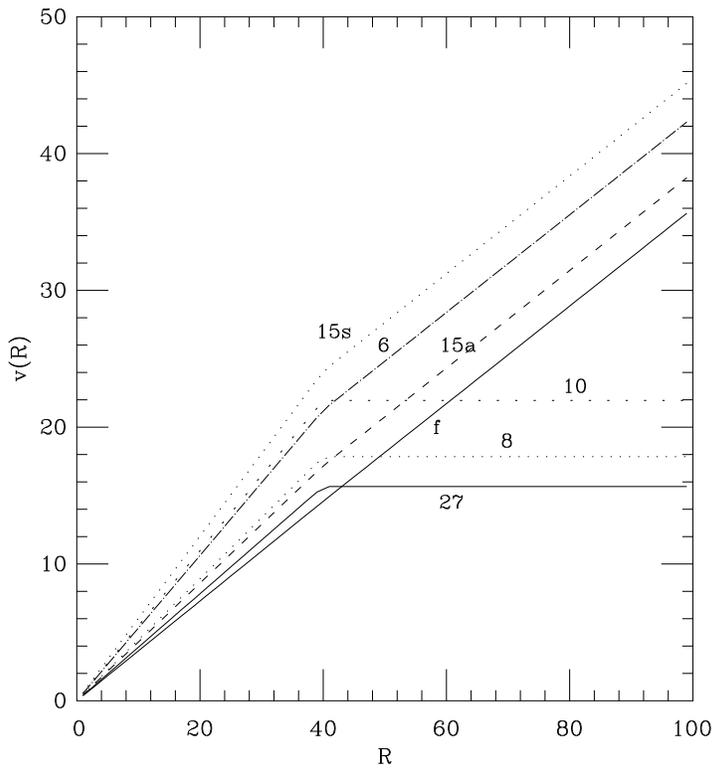}
\caption{Static sources potential using a density distribution proportional
to delta function which is zero every where except at the two points where the
vortex first enters and exits the Wilson loop.}
\label{delta}
\end{figure}

Qualitative agreement with Casimir scaling is observed for all the
axially symmetric distributions I tried. This
is true even if one defines the density to be zero everywhere except on the 
outer boundary ($r=a$) of the vortex. Fig.\ \ref{Pdens} plots potentials for this
distribution with $a=20$ and $f=0.1$. Fig.\ \ref{delta} shows potentials for 
the maximally non-axially symmetric core where the flux is zero everywhere 
except at the two points where
the vortex first enters and exits the Wilson loop. A linear regime still
exists at intermediate distances but qualitative agreement with Casimir scaling
is lost: slope of potentials with larger Casimir operators have smaller string
tensions. For example, the string tension for representations 6 and 8 are 
larger than the ones for representations 15-antisymmetric and 27. 
In this case the order of potentials at long distances 
changes as well. For example, the potential for representation 27 is less than
the potential for representation 8 in the screened regime. 
It still remains true, however, that the zero triality representations screen, 
and non-zero triality ones approach the fundamental string tension at long
distances.  However the non-convexity of potentials is almost gone.
Note that this flux distribution is probably unphysical. We expect the flux
in the lowest energy vortex to be axially symmetric. The fact that 
potentials obtained from physical flux distributions agree qualitatively with 
Casimir scaling, is the strength of the vortex model. However recent numerical 
simulation results \cite{Bali99} show that potentials are quantitatively in agreement 
with Casimir scaling with an accuracy of 5 percent, the feature that is lost in 
fat-center-vortices results. For example, from Fig.\ ref{Pdens}, ratios of 
potentials for representations 6, 8, 10, 15-symmetric, 15-antisymmetric, and
27 to the fundamental representation are 2, 1.82, 3, 5.4, 2.7, and 3.4,   
respectively whereas the same ratios for Casimir scaling would be 2.5, 2.25, 
4.5, 7, 4, and 6, respectively. 

One can get similar results for potentials at intermediate and long distances
using one dimensional functions for $\alpha(x)$ similar to those of
ref.\ \cite{Fabe97}. I have tried several functions. An example is:
\begin{equation}
\beta(x_{v})= \left \{ \begin{array}{lll} 
\frac{4\pi}{\sqrt{3}} & \mbox{$x_{v}>a$ } \\ 
0 & \mbox{$x_{v}<-a$ }\\ 
\frac{2\pi}{\sqrt{3}}+\frac{2\pi}{\sqrt{3}}(\exp\{b[1-\frac{1}{(x_{v}/a+1)^2}]\}- \exp\{b[1-\frac{
1}{(x_{v}/a-1)^2}]\}) &  \mbox{$-a<x_{v}<a.$}
\end{array}
\right.
\end{equation}
$\alpha_{R}(x_{v})$ is then given by Eqn.\ \ref{alpha-def}.
Note that in this case no density distribution is defined and therefore no 
integration is needed to find $\beta(x)$.
For moderate $b$ this results in a flux profile similar to the one used in 
ref.\ \cite{Fabe97} and leads to qualitative Casimir scaling. 
However for $b$ large enough this gives similar plot to Fig.\ \ref{delta}
and violates Casimir scaling. 
Thus integrating a circularly symmetric distribution inside the Wilson loop
is more physical than just picking a function $\beta(x)$ at random.
An arbitrary one dimensional distribution is not necessary consistent
with axial symmetry, and in the above example, large $b$ certainly does
not correspond to any axially symmetric distribution.

\section{Conclusion}

By applying the fat-center-vortices model to $SU(3)$ and using presumably 
physical axially symmetric density distributions for the vortex, I
showed that for several representations there exists a region at intermediate 
distances in which the
static potential is linear and qualitatively in agreement with Casimir
scaling. This is also in agreement with the observation in SU(3) 
simulations of a linear potential in proportion to Casimir ratio of the 
representation \cite{Camp86,Mich92,Mich98,Deld98,Bali99}. However, the Casimir proportionality
is dependent on the flux distribution in the vortex 
and it is possible to lose this feature by changing 
the distribution function to a non-axially symmetric distribution. At large 
distances, zero-triality representations will be screened and the potentials 
for non-zero triality representations parallel the one for the 
fundamental representation. Some of the expected features of the screening
pattern are also lost for non-axially symmetric distributions.
The conclusion is that Casimir scaling and the pattern of screening depend
on the detailed vortex structure and are not simple kinematic 
consequence of the fat center vortex picture. However it is also clear
that these properties are rather robust and are likely  to survive with 
physically realistic vortices. On the other, potentials are not 
quantitatively in agreement with Casimir scaling as predicted by recent numerical
simulations. This suggests that the fat-center-vortices model needs further
refinement if it is to remain viable. In particular, one may need an appropriate
physical flux distribution of vortex sizes. Further numerical studies of these
issues is in progress.

\section{Acknowledgement}

I wish to thank Claude Bernard for his help in
this work.

\clearpage
\vspace*{100pt}

\clearpage
\vspace*{100pt}

\clearpage
\vspace*{100pt}

\end{document}